# Hybrid reciprocal lattice: application to layer stress appointment in GaAlN/GaN(0001) systems with patterned substrates

Jarosław Z. Domagała,[a] Sérgio L. Morelhão,[b,*] Marcin Sarzyński,[c] Marcin Maździarz,[d] Paweł Dłużewski,[d] and Michał Leszczyński [c,e]

[a] Institute of Physics, Polish Academy of Sciences, Al. Lotnikow 32/46, 06-668 Warszawa, Poland.
[b] Institute of Physics, University of Sao Paulo, C.P. 66318, CEP 05315-970, São Paulo, SP, Brazil.
[c] Institute of High Pressure Physics (UNIPRESS), Polish Academy of Sciences, Sokolowska 29/37, 01-142 Warszawa, Poland.
[d] Institute of Fundamental Technological Research (IPPT PAN), Pawińskiego 5 B, 02-106 Warszawa, Poland.
[e] TopGaN Ltd, Sokolowska 29/37, 01-142 Warszawa, Poland.

**Abstract**

Epitaxy of semiconductors is a process of tremendous importance in applied science and optoelectronic industry. Controlling of defects introduced during epitaxial growth is a key point in manufacturing devices of high efficiency and durability. In this work, we demonstrate how useful hybrid reflections are on the study of epitaxial structures with anisotropic strain gradients due to patterned substrates. High accuracy to detect and distinguish elastic and plastic relaxations are one of the greatest advantages of measuring this type of reflection, as well as the fact that it can be exploited in symmetrical reflection geometry on a commercial high-resolution diffractometer.

INTRODUCTION

X-ray diffraction is a well established technique for structural analysis of electronic and optoelectronic devices based on single crystal substrates and epitaxial layers. However, advances in materials and processing technologies as well as in X-ray instrumentation (sources, focusing optics, and detectors) are creating a demand for new approaches on how to further exploit this fantastic tool of structural analysis. One example of a material is the gallium nitrite, a material that has revolutionized the optoelectronic industry in the early 90s, providing blue photons out of a light-emitting diode.[1] But, only very recently it could be synthesized in the form of high quality single crystals opening new opportunities in manufacturing electronic and optoelectronic devices.[2-6] Controlling of defects in III-nitride semiconductor devices is an important issue in terms of processing technologies,[7-10] and it has also to be addressed when using bulk GaN.

High-flux X-ray beam impinging on perfect single crystals, such as the substrate material used in optoelectronic devices, gives rise to intensity measures further way from the ordinary Bragg peaks. Although most of these intensities are caused by diffractions of higher order,[11-12] they can also be related to diffuse scattering.[13] Since compact high-flux sources and zero noise X-ray detectors are becoming affordable as home lab equipments, intentional or accidental observations of such intensity features are expected to become more often. Being able to properly interpret intensity measures of this kind is crucial at the actual state of our technological development in both research fields, new materials and X-ray metrology.

Occurrence of X-ray re-scattering in epitaxial-layer/substrate systems, named hybrid reflections, is a phenomenon that has been exploited in cubic (001) crystals only,[14-16] althouth already observed in hexagonal systems: the first observation of hybrids in non-cubic materials was reported by Domagala in 2010 for GaN layer on SiC(0001).[17] Hybrid reflections have origin in multiple diffraction cases,[18] which can be indexed in bulk materials by available softwares.[19-21] In III-nitredes, investigation of multiple diffraction were carried out recently.[22-23] For strain analysis, hybrid reflections offer the benefit of obtaining structural 3D information in diffraction geometries well suitable for most standard high-resolution diffractometers with laboratorial (compact) X-ray sources. However, the 3D nature of this phenomenon requires a high level of expertise in single crystal multiple diffraction geometries and in both reciprocal and real space. In this work, we present a general approach to facilitate exploration of hybrid reflections in non-cubic systems, and we use it to elucidate features in reciprocal space owing to strain in AlGaN epitaxic-layers grown by MBE on highly perfect GaN(0001) patterned substrates.

**Bragg cone lines**. In real space, excitement of Bragg reflections is ruled by the relationship $\boldsymbol{k}\cdot\boldsymbol{G}=-|\boldsymbol{G}|^2/2$ between the incident wavevector $\boldsymbol{k}$ and a diffraction vector $\boldsymbol{G}$, which gives rise to the Bragg cone (BC) representation of the condition of diffraction illustrated in Figure 1a. When analyzing single crystals of one large face, as the substrate material employed for epitaxial growth, it is convenient to choose a reference system for describing the direction of the wavevector $\boldsymbol{k}=-(2\pi/\lambda)[\cos\omega\cos\varphi,\cos\omega\sin\varphi,\sin\omega]$, in terms of instrumental angles such as the incidence angle $\omega$ and azimuth $\varphi$. By projecting the diffraction vector in this reference system, $\boldsymbol{G}=|\boldsymbol{G}|[\sin\gamma_G\cos\alpha_G,\sin\gamma_G\sin\alpha_G,\cos\gamma_G]$ with $|\boldsymbol{G}|=(4\pi/\lambda)\sin\theta_G$, the BC equation can be replaced by the relationship

$$\cos(\varphi-\alpha_G)=\frac{\sin\theta_G-\sin\omega\cos\gamma_G}{\cos\omega\sin\gamma_G} \qquad (1)$$

to be fulfilled by the instrumental angles for Bragg diffraction takes place.

This relationship, Eq. 1, provides the simplest way to collapse into a 2D representation the 3D geometry of multiple diffraction in single crystals. It allows an easy understanding of all reflections that can be excited when mapping one particular region of the $\omega$ and $\varphi$ angular space.

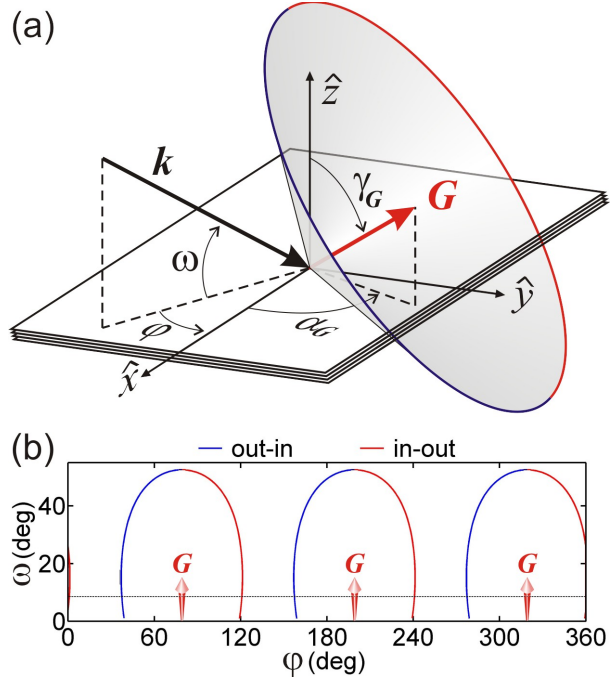

**Figure 1**. (a) Reference system used to describe the incident wavevector ***k*** and the Bragg cone (BC) of a diffraction vector ***G***. (b) Excitation conditions in terms of $\omega$ and $\varphi$ angles for equivalent reflections around a 3-fold symmetry axis. Curved (blue/red) and straight (dashed) lines stand for BCs of asymmetric and symmetric reflections, respectively.

For instance, Figure 1b shows the case of an azimuthal axis with 3-fold symmetry, such as the [0001] axis in hexagonal crystals. Curved lines stand for BCs of asymmetrical reflections, while diffraction vectors along the azimuthal axis have their BCs represented as straight lines in the $\omega{:}\varphi$ graphs (Appendix A), as the dashed line in Figure 1b.

In single crystals, observation of multiple BC lines with monochromatic and collimated X-ray beam is possible near their intersections owing to dynamical coupling between the diffracted waves of each reflection.[24,25] With high flux sources (synchrotron and microfocus with focusing optics), even BC lines of forbidden reflections are turned on in a few milliradians range from the intersection point. A benchmark demonstration of this long-range dynamical coupling is the intersection of BC lines from 002 and 111 reflections in silicon (001) crystals, Figure 2a. Rocking curves of the symmetrical reflection near this intersection point could present two distinct peaks instead of none expected for the forbidden 002 Si reflection, as seen in Figure 2b. The angle and position of the BC lines predicted by Eq. 1 is given in Figure 2c. Monitoring the intensity distribution over BC lines as well as their angle of intersection have been used to access surface defects and residual strain in nanostructured devices grown on semiconductor substrates.[24,26] For instance, enhancement of intensity along the 111 BC line above the 002 reflection, Δω>0 in Figure 2a, is related to the presence of surface defects that increase the re-scattering of diffracted beam when the 111 reflection is excited, such as reported in GaAs (001) substrates.[27] Moreover, BC lines have also been seen decorating X-ray images collected by zero-noise detectors, in a process called multiple diffuse scattering.[13]

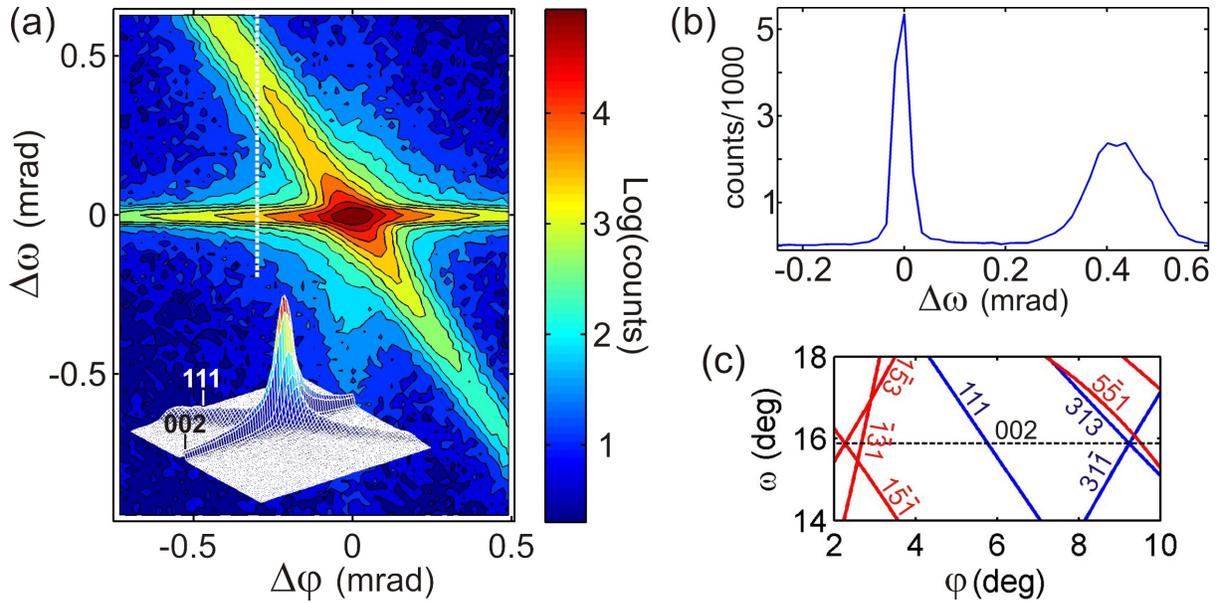

**Figure 2**. Intersection of Bragg cones in silicon (001). (a) Two-dimensional intensity profile of the forbidden 002 reflection at near intersection point with the 111 reflection, after ref. 29. X-rays of 8.34 keV, beam divergences < 25µrad (both axes). (b) Rocking-curve (ω-scan) of the forbidden reflection showing two peaks at azimuth Δφ = -0.3 mrad, dashed-white line in (a). (c) Nearby BC lines according to Eq. 1.

**Hybrid reflections in hexagonal lattices with strain gradients**. In epilayer/substrate systems, coupling of diffracted waves can be observed further way from intersection points of BC lines due to the re-scattering of diffracted beams by another material with different lattice parameters. Angular conditions for exciting such re-scattering events fall over or very near BC lines of the substrate since, in general, these systems present small lattice mismatch. A simple ω:φ graph from Eq. 1 for the substrate is then capable of providing most of the necessary information to locate and to use hybrid reflections for stress analysis. After selecting a region of the angular space to probe for re-scattering events, the reciprocal space must be mapped to discriminate location and shape of hybrid RLPs (reciprocal lattice points) that might exist there. Their locations in the vicinity of a substrate RLP are given by [15]

$$\Delta \boldsymbol{P} = h' \Delta \boldsymbol{a}^* + k' \Delta \boldsymbol{b}^* + l' \Delta \boldsymbol{c}^*. \qquad (2)$$

$h'$, $k'$, and $l'$ are integer numbers, and $\Delta \boldsymbol{g}^* = \boldsymbol{g}_L^* - \boldsymbol{g}_S^*$ (with $\boldsymbol{g}^* = \boldsymbol{a}^*$, $\boldsymbol{b}^*$ or $\boldsymbol{c}^*$ as the edges vectors of the reciprocal unit cell) characterize variation in the reciprocal lattice between both materials where subindexes $L$ and $S$ stand for epilayer and substrate lattices, respectively. In terms of longitudinal $Q_z$ and transverse $Q_{xy}$ coordinates in reciprocal spaces maps, the location of the hybrid RLP are

$$Q_z = \Delta \boldsymbol{P} \cdot \hat{\boldsymbol{z}} \text{ and } Q_{xy} = \Delta \boldsymbol{P} \cdot \hat{\boldsymbol{k}}_{xy} \qquad (3)$$

where $\hat{\boldsymbol{k}}_{xy}$ is the in-plane direction of the incident wavevector, i.e. $\hat{\boldsymbol{k}}_{xy} = -[\cos\varphi, \sin\varphi, 0]$ according to Figure 1.

Since hybrid RLPs are originated in the differences between both lattices, their shapes are strongly affected by strain gradients, a fact that had been predicted, but had not been observed yet. On the other hand, in totally relaxed epilayers (absence of strain gradients), the shapes of hybrid RLPs are susceptible to the ratio between in-plane and out-of-plane mosaicity.[15]

Gradient and anisotropy of strain can be introduced by grooves in epilayer/substrate materials. Grooves with flat terrace and along the *y* direction, e.g. Figure 3a, gives rise to a nearly constant $\varepsilon_y$ strain, while the $\varepsilon_x$ strain may vary from maximum values in the center of the terraces towards minimum values in the edges. In hexagonal systems with (0001) surface and grooves along the $[\bar{1}10]$ direction, the anisotropy of strain also implies in unit cells with distorted basal plane where the γ angle differs from 120° by Δγ, Figure 3b. In terms of the in-plane strain components, strained epilayer unit cells are described by

$$\frac{\Delta a}{a} = \frac{\Delta b}{b} = \frac{\varepsilon_x + \varepsilon_y}{2} + \cos\gamma\, \frac{\varepsilon_x - \varepsilon_y}{2}, \quad \frac{\Delta c}{c} = -\frac{C_{13}}{C_{33}}(\varepsilon_x + \varepsilon_y), \text{ and } \Delta\gamma = -\sin\gamma(\varepsilon_x - \varepsilon_y).$$

Elastic strain relaxation—without misfit dislocations—at groove terraces can be accompanied by a deflection of the lattices, causing misorientation of diffraction vectors at near the groove edges. A pictorial scheme accounting for both effects, strain gradient and misorientation of relaxed epilayer regions, are shown in Figure 4. By mapping the reciprocal space around the 001 substrate RLP with the sample at a proper azimuth, hybrids such as $103_S + \bar{1}0\bar{2}_L$ and $10\bar{2}_L + \bar{1}03_S$ will have shapes reveling the presence of these two effects.

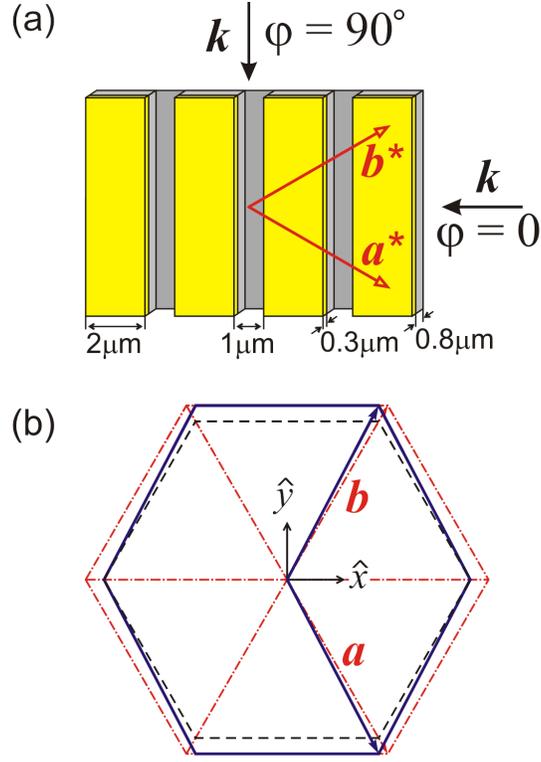

**Figure 3**. (a) AlGaN/GaN(001) sample with grooves along the $[\bar{1}10]$ crystallographic direction. (b) Basal plane of the epilayer's unit cell when fully strained (dash-dot line), relaxed (dashed line) or partially relaxed (solid line) along the x direction only.

To easily access in-plane strain values from a measure of $Q_{xy}$ either in reciprocal space maps (RSMs) or in transverse scans, Eq. 3 is rewritten as a function of in-plane lattice mismatches along the x and y directions, i.e. $\varepsilon_x^s = (\Delta d/d)_{[110]}$ and $\varepsilon_y^s = (\Delta d/d)_{[\bar{1}10]}$ respectively. It results in

$$Q_{xy} = \frac{2\pi}{a_s}(h'+k')\cos\varphi\,\varepsilon_x^s + \frac{2\pi}{a_s\sqrt{3}}(-h'+k')\sin\varphi\,\varepsilon_y^s \qquad (4)$$

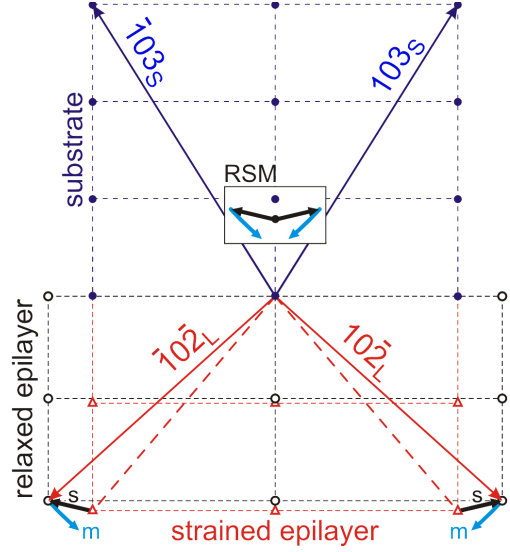

**Figure 4**. Scheme of how strain gradient (arrows s) and misorientation (arrows m) of epilayer's diffraction vectors are seen when mapping hybrid reflections in reciprocal space.

that can then be assigned to the epilayer strain components by means of $a_s(1+\varepsilon_{x,y}^s)=a_{0,L}(1+\varepsilon_{x,y})$ where $a_s$ and $a_{0,L}$ are the unstrained substrate and epilayer in-plane lattice parameters.

EXPERIMENTAL SECTION

**Sample preparation**. AlGaN epilayer were deposited by Molecular Beam Epitaxy (MBE) on hexagonal GaN (0001) substrate.[3] Nominal epilayer thickness 0.3 μm. Lithograph were carried out after growth to produce square grooves as indicated in Figure 3a. Here, Bragg reflections are indicated by *hkl* indexes regarding the primitive unit cell, e.g. Figure 3b.

**X-ray measurements**. X-ray data have been collected on a Philips X'Pert-MRD high-resolution diffractometer with Cu tube and proportional gas detector. Incident beam optics: X-ray mirror; four-crystal asymmetric 220 Ge monochromator ($\Delta\lambda/\lambda = 5\times 10^{-5}$); beam divergences 0.005° and 2° in the incidence plane and axial direction, respectively; beam size at sample position 0.28 mm × 1 mm. Diffracted beam optics: three-bounce 220 Ge analyzer crystal for triple axis or 1/4° slit for double axis diffractometry. Extra features: diffraction vector of symmetrical reflections were precisely aligned with the φ rotation axis by a mechanical device assembled to the standard sample holder of the diffractometer.

RESULTS AND DISCUSSIONS

**Logitudinal scans**. One-dimensional reciprocal space scans along the *c*-axis, i.e. 2θ/ω scans, carried out as a function of sample's azimuth reveal hybrid RLPs with index $l'$ = -2, -1, 0, and +1, Figure 5a. $h'k'l'$ indexes of the hybrids are inferred from the nearby BC lines shown in Figure 5b. BC lines of indexes $hkl$ where $l \leq 0$ gives rise to hybrids of indexes $h'k'l' = hkl$, while $h'=-h$, $k'=-k$, and $l'=1-l$ are the hybrid indexes for BC lines where $l \geq 1$. For instance, $10\bar{2}$ and $\bar{1}0\bar{2}$ are hybrid indexes at φ = 10° in Figure 5a, and they are excited by $10\bar{2}$ and 103 BC lines, respectively. In other words, hybrids have the same indexes of the epilayer reflections taking part in the re-scattering events, such as in the illustration in Figure 4. $l'$ = 0 are hybrids coincidental with *n*-beam diffraction of the substrate and $l'$ = +1 are the epilayer *n*-beam diffractions. Therefore, actual hybrids revealed by the 2θ/ω scans in Figure 5a are those of $l'$ = -2 and -1. Their positions along the *c*-axis are explained by a perpendicular lattice mismatch $\varepsilon_z^s = (c_L - c_S)/c_S \simeq -7.1 \times 10^{-3}$, which follows from

$$2\Delta c^* = 2\left(\frac{2\pi}{c_L} - \frac{2\pi}{c_S}\right) \simeq (4\pi/\lambda)[\sin(17.097/2) - \sin(16.850/2)] = 0.0174 \text{ Å}^{-1}.$$

According to Eqs. 2 and 3, hybrids with non zero $h'$ and $k'$ indexes can be seen on one-dimensional scans along the *c*-axis in two situations: *i)* when there are fully strained epilayer regions in which $\Delta a^* = \Delta b^* = 0$; or *ii)* when the projection of $\Delta P$ along $\hat{k}_{xy}$ is null. Since the hybrids in Figure 5a are seen at different azimuths from φ = 10° to 50°, the in-plane lattice mismatches are null or very small as can be checked by applying Eq. 4 in the observed hybrids. A direct determination of epilayer composition is therefore possible in such case. By using tabulated values for the GaN and AlN compounds (Appendix A), we have that in fully strained Al$_x$Ga$_{1-x}$N epilayer on GaN (001) $x \approx -19.4 \varepsilon_z^s$, and the above value of lattice mismatch leads to $x = 0.138 \pm 0.004$.

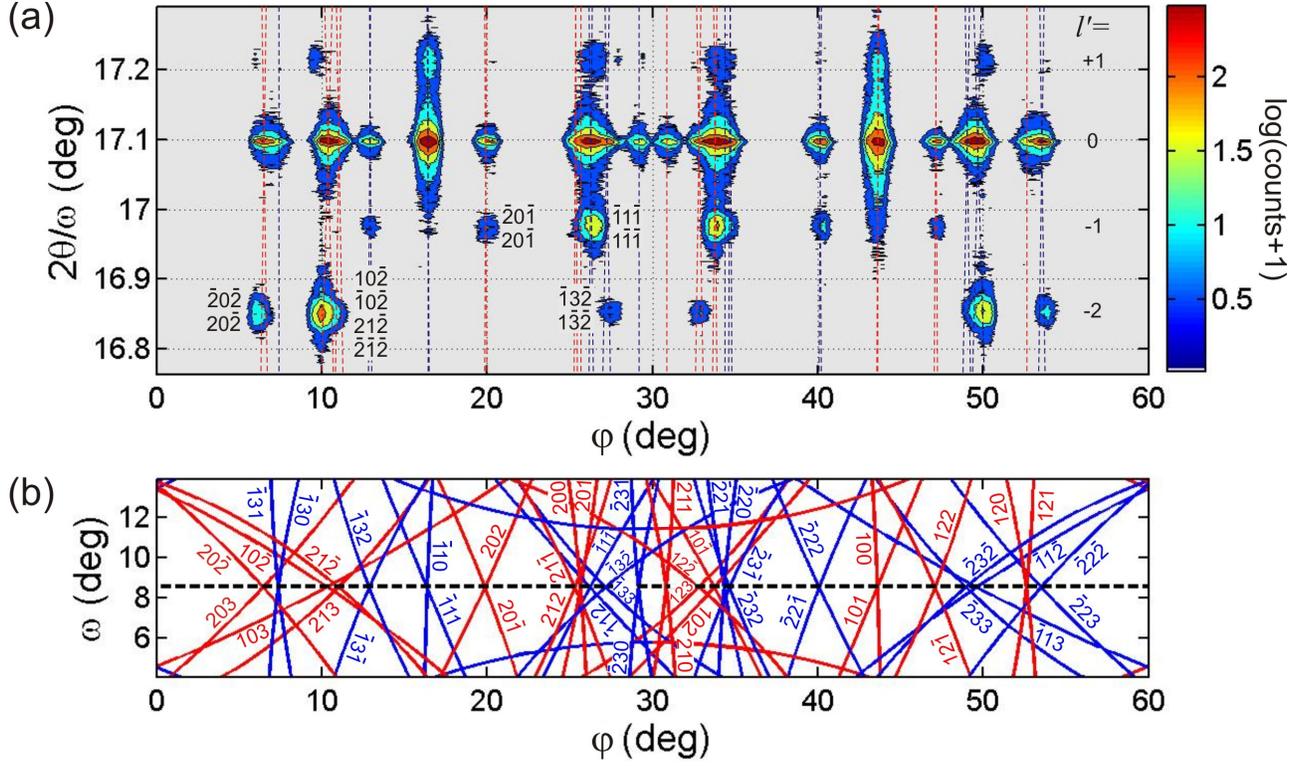

**Figure 5**. (a) 2θ/ω scan of the forbidden 001 GaN reflection as a function of sample's azimuth. Reference direction [110] (φ=0). *h′k′l′* indexes are explained in the text. (b) BC lines in GaN (001), also shown in (a) as dashed lines after proper re-scaling.

**Reciprocal space mapping**. A RSM carried out at φ = 10° is presented in Figure 6a. The strongest hybrid intensity occurs at $Q_{xy}$ = 0, evidencing that most of the epilayer regions fully strained (no in-plane lattice mismatch) with respect to the substrate lattice taking part in the re-scattering events. However, the pattern formed by the weak intensity distribution around the hybrid spot closely resembles the expected pattern schematized in Figure 4 for the case of gradual release of strain and misorientation of relaxed epilayer regions. A similar pattern is also seen in Figure 6b where the RSM carried out at φ = 34° encompasses hybrids with index *l'* = -1. RSMs in Figures 6c,d measure equivalent reflections, but at azimuths increased by 60° and hence closer to the *y* direction. The patterns of weak intensity in these latter maps are observed to shrink according to cosφ regarding those in Figures 6a,b. It agrees with a relaxation of the epilayer preferentially along direction [110] perpendicular to the grooves.

From one RSM to the other, modification in the two-dimensional intensity profile of the substrate 001 *n*-beam diffraction spots can also be noticed. The streaks of weak intensity of the substrate are nearly parallel to those around the hybrid spots—all streaks are pointed out by arrows in Figures 6a,b. It confirms that both lattices have deviated from flatness at grooves terraces, indicating that the relaxation caused by the grooves have an elastic character.

Along the grooves some misfit dislocations (plastic relaxation) may have been introduced in the epilayer, judging by the small split of the hybrid spots that can be seen in Figures 6c,d. For a split of $(0.57\pm0.01)\times10^{-3}$ Å$^{-1}$ between the hybrids $01\bar{1}$ and $0\bar{1}\bar{1}$ at φ = 94°, Figure 6d, Eq. 4 provides a lattice mismatch $\varepsilon_y^s = -(2.4\pm0.2)\times10^{-4}$ along the grooves. It corresponds to an epilayer relaxation of about 7.3±0.6 % when taking $a_s$ = 3.1885 Å and $a_{0,L}$ = 3.178 Å

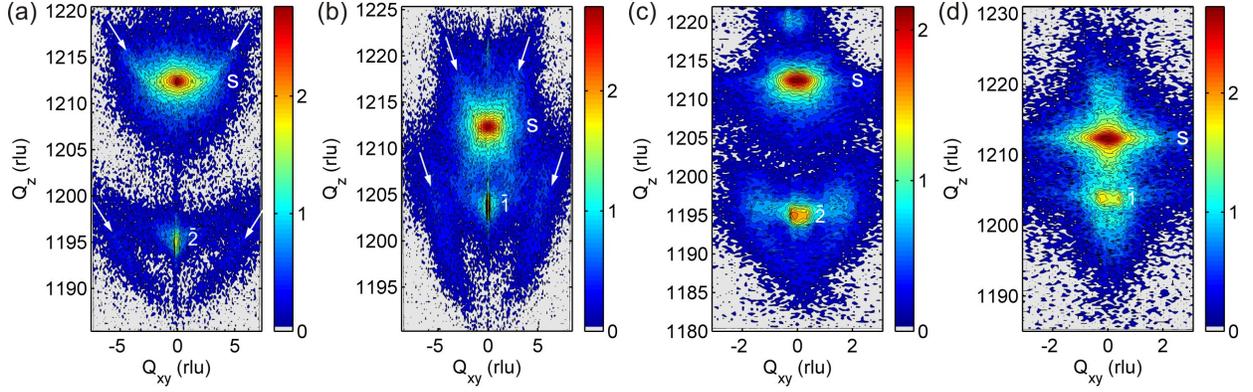

**Figure 6**. Reciprocal space maps of forbidden 001 reflection in AlGaN/GaN (001) at different azimuths: (a) φ=10°, (b) φ=34°, (c) φ=70°, and (d) φ=94°. Substrate n-beam diffraction spots are labeled S, while hybrid spots are labeled by their *l'* index, $\bar{1}$ or $\bar{2}$. Streaks of week intensity mentioned in the text are indicated by arrows. Color bars are log(counts+1). Rec. latt. units (rlu) = 0.001 Å$^{-1}$.

(composition *x* = 0.138).

For the sake of comparison, the usual procedure for strain analysis in epilayer/substrate systems have also been carried out. RSMs of symmetrical and asymmetrical reflections at azimuths perpendicular (φ = 0) and parallel (φ = 90°) to the grooves are shown in Figure 7. The correlation between deflection and strain of both lattices mostly along the *x* direction is clearly noticed in Figures 7a,b. That is the behavior expected from finite element analysis (Appendix B). Asymmetric $\bar{1}\bar{1}4$ reflections in Figures 7c,d evidence epilayer relaxation perpendicular to the grooves, while the asymmetric $1\bar{1}5$ reflections in Figures 7e,f show a fully strained epilayer along the grooves. An in-plane relaxation of 50% implies in a perpendicular lattice mismatch of $\varepsilon_z^s = -6.2 \times 10^{-3}$ (composition *x* = 0.138, see Appendix A), and that is the experimental value obtained by a direct measure of the spot-to-spot distance along $Q_z$ in Figure 7a, i.e.

$$\varepsilon_z^s = -(2439.1 - 2423.7)/2423.7 = -(6.4 \pm 0.2) \times 10^{-3}.$$

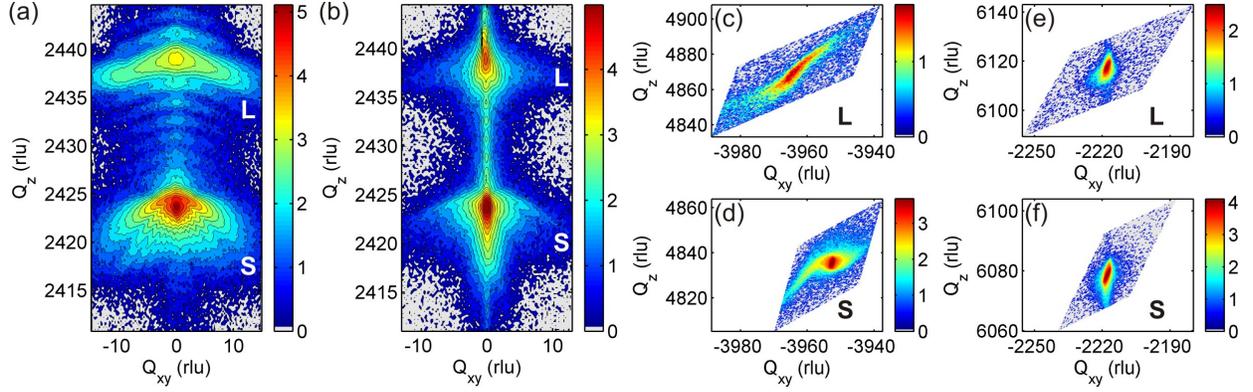

**Figure 7**. Reciprocal space maps of different reflections in AlGaN/GaN (001): (a) 002 at $\varphi = 0$; (b) 002 at $\varphi = 90°$; (c,d) $\bar{1}\bar{1}4$, $\varphi = 0°$; (e,f) $1\bar{1}5$, $\varphi = 90°$. Labels L and S stand for epilayer and substrate reflections. Color bars are log(counts+1). Rec. latt. units (rlu) = 0.001 Å$^{-1}$.

In-plane lattice mismatches accessed via asymmetric reflections, Figures 7c-f, have an accuracy limit not smaller than $5 \times 10^{-4}$. It is imposed by the fact that the diffraction spots are relatively broad regarding their separation in reciprocal space. Then, asymmetric reflections in Al$_{0.14}$Ga$_{0.86}$N/GaN(001) have been unable to detect epilayer relaxation smaller than 15%. Hybrid reflections, on the contrary, have shown an accuracy of $2 \times 10^{-5}$ in detecting in-plane lattice mismatch, and hence the capability to probe variations of 1% in the epilayer relaxation. Such accuracy is corroborated by the fact that hybrid reflections are re-scattering processes through the epilayer/substrate interface. Elastic relaxation of lattice mismatch through the interface takes place in a much smooth length scale than an abrupt plastic deformation caused by misfit dislocations. Therefore, the sharp splits in $Q_{xy}$ of the hybrid spots seen in Figures 6c,d are strong evidences of plastic relaxation, although in an amount that is hard to detect by standard measures of asymmetric reflections even when detailed X-ray dynamical diffraction simulation is carried on.[28]

CONCLUSION

Patterned GaN (0001) substrate was very successful in promoting the elastic relaxation of a 0.3 mm thick $Al_{0.14}Ga_{0.86}N$ epilayer in the direction perpendicular to the lithographed stripes, preventing misfit dislocations with Burgers vector perpendicular to the stripes in the structure. However, along the stripes, a relaxation of about 7% could be credited to the presence of misfit dislocations. The X-ray diffraction method demonstrated and used in this work exploits 2nd-order scattering events at the epilayer/substrate interface, which allows detection of in-plane lattice mismatches with an accuracy that is at least 10 times better than measuring single asymmetrical reflections of the layer and substrate lattices separately. Direct determination of epilayer composition by simple 2θ/ω scans in symmetrical reflection geometry is also possible. Moreover, the proposed general approach to locate and index such re-scattering events greatly facilitated the implementation of the method in hexagonal crystal structures.

APPENDIX A

**Two-dimensional Bragg cone representation**. To project the crystal's reciprocal space on the reference system used to describe the instrumental ω and φ angles, $A = H\boldsymbol{a}^* + K\boldsymbol{b}^* + L\boldsymbol{c}^*$ is taken as the reciprocal vector aligned along the φ rotation axis, and $\boldsymbol{B} = B_1\boldsymbol{a}^* + B_2\boldsymbol{b}^* + B_3\boldsymbol{c}^*$ a direction in reciprocal space that lies in the incidence plane (pointing upstream) when φ = 0, then $\hat{\boldsymbol{x}} = \hat{\boldsymbol{y}} \times \hat{\boldsymbol{z}}$, $\hat{\boldsymbol{y}} = \hat{\boldsymbol{z}} \times \boldsymbol{B}/|\boldsymbol{B}|$, and $\hat{\boldsymbol{z}} = \boldsymbol{A}/|\boldsymbol{A}|$. From a full list of reflection indexes with no null structure factors,[34] the diffraction vectors $\boldsymbol{G}$ are taken and their polar and azimuthal angles calculated as $\gamma_G = \cos^{-1}(\hat{z} \cdot \boldsymbol{G}/|\boldsymbol{G}|)$ and $\alpha_G = \tan^{-1}(\hat{y} \cdot \boldsymbol{G}/\hat{x} \cdot \boldsymbol{G})$, respectively; $\alpha_G$ must be accounted in the four quadrants. Eq. 1 provides two solutions for the azimuth φ as a function of the incidence angle ω, which are plotted as the out-in and in-out BC lines in the ω:φ graphs, e.g. Figures 1b, 2c, and 5b.

|      | $a$ (Å) | $c$ (Å) | $C_{13}$ | $C_{33}$ |
|------|---------|---------|----------|----------|
| GaN  | 3.1885  | 5.1850  | 103      | 405      |
| AlN  | 3.112   | 4.982   | 108      | 373      |

Table A1. Lattice and elastic constants of AlN and GaN compounds.[30-33]

**Strain calculation in Al$_x$Ga$_{1-x}$N alloy**. By applying Vegard's law to AlN and GaN lattice and elastic constants in Table A1, the perpendicular lattice mismatch of Al$_x$Ga$_{1-x}$N epilayer on GaN (0001) as a function of composition $x$ and average in-plane relaxation $\langle R \rangle$ has a linear behavior that, for $x < 0.2$ (Figure A1), can be summarized as

$$\varepsilon_z^s = (\Delta c/c) = -[5.16 - 1.24 \langle R \rangle] \times 10^{-2} x . \qquad (A1)$$

For instance, an epilayer with composition $x = 0.138$ would have $\varepsilon_z^s = -7.1 \times 10^{-3}$ when fully strained, $\langle R \rangle = 0$, and $\varepsilon_z^s = -6.2 \times 10^{-3}$ when totally relaxed along one in-plane direction, i.e. $\langle R \rangle = 0.5$.

APPENDIX B

**Elastic mismatch accommodation in AlGaN/GaN interface by finite element calculation**. A nonlinear finite element problem was solved with the use of a chemo-hyperelastic user element programmed in FEAP program.[35-37] The anisotropic hookian model used is based on Hencky measure of lattice strain; the chemo-elastic coupling is modelled with the use of Vegard's law. In this approach a 3D displacement field is spanned on 27-nodes (brick element) by means of 2-nd order Lagrangean shape functions. Only 8 corner nodes are used for modelling the aluminium fraction field stored in nodes on the 4-th freedom degree and spanned by means of 1-st order Lagrangian shape functions. Such a continuous distribution of elastic and chemical fields was essential mainly for modelling the residual stresses in the interfacial zone. The zone AlGaN/GaN was discretized by means of a very narrow single layer of finite elements thickness 10nm, see Figure B1. Third three deformation freedoms in nodes corresponded to displacements $x$, $y$, and $z$; the 4-th degree was used to input the field of aluminum molar fraction. The plane strain boundary condition in the $y$ direction was imposed for all nodes of the mesh. Due to the symmetry the displacements in the x direction have been fixed on external surfaces in the $x$ directions. At the bottom, the mesh was fixed in the $z$ direction.

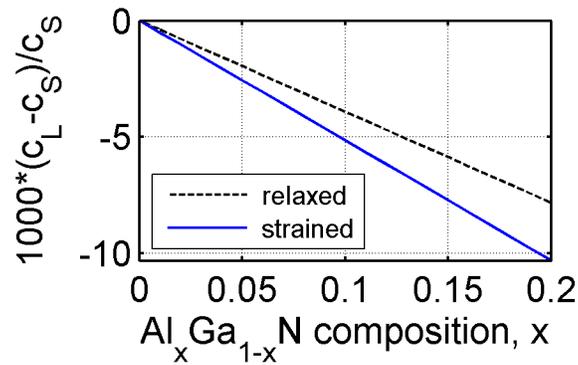

**Figure A1**. Perpendicular lattice mismatch in relaxed and fully strained $Al_xGa_{1-x}N$ on GaN (0001) as a function of composition $x$.

AUTHOR INFORMATION

**Corresponding Author**

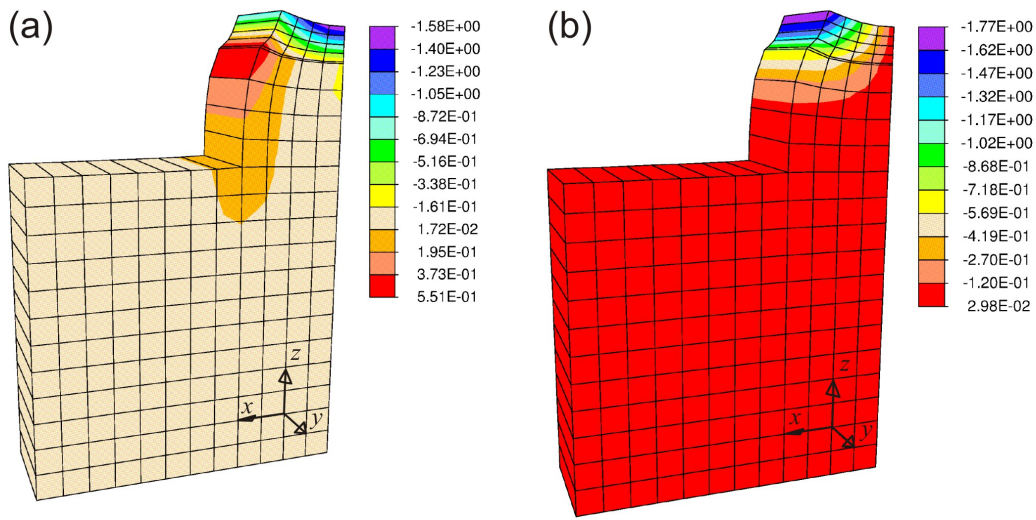

**Figure B1**. Finite element analysis. (a,b) Deformation with respect to GaN bulk crystal, (a) along *z*, and (b) along *x*. Mesh view with 100 times enlarged displacements to better visualise the character of lattice distortion.

*E-mail: morelhao@if.usp.br.

ACKNOWLEDGMENTS

The authors acknowledge support in part by the project WISEGaN sponsored by the polish National Centre for Research and Development (NCBiR) in the frame of Japan-V4 program, N N519 647640 founded by the Polish Ministry of Science and Higher Education. S.L.M. also acknowledge the Brazilian founding agency CNPq (grants 306982/2012-9 and 452031/20150).